%% file: main_spl.tex
\newif\ifextended
\newif\ifprov
\extendedtrue     
\provfalse         
\newcommand{\pathStyleFiles}{./}
\newcommand{\pathBibFilesA}{./biblio}
\documentclass[journal]{\pathStyleFiles/IEEEtran}
\input{\pathStyleFiles/defSPL.tex}

\input{\pathStyleFiles/defCdes.tex}
\input{\pathStyleFiles/defGras.tex}
\title{Unsupervised linear component analysis \\ for a class of probability mixture models}
\author{Marc Castella, {\em Senior Member, IEEE}.  \thanks{M. Castella (corresponding
    author) is with SAMOVAR, T{\'e}l{\'e}com SudParis, Institut Polytechnique de
    Paris, 91120 Palaiseau, France. \texttt{marc.castella@telecom-sudparis.eu}  } }
\floatname{algorithm}{Alg.}

\ifprov
\renewcommand{\modifA}[1]{#1}   
\else
\renewcommand{\temp}[1]{}
\renewcommand{\modifA}[1]{#1}
\renewcommand{\modifB}[1]{#1}
\fi

\begin{document}
\maketitle

\begin{abstract}
  We deal with a model where a set of observations is obtained by a linear
  superposition of unknown components called sources. The problem consists in
  recovering the sources without knowing the linear transform. We extend the
  well-known Independent Component Analysis (ICA) methodology. Instead of
  assuming independent source components, we assume that the source vector is a
  probability mixture of two distributions. Only one distribution satisfies the
  ICA assumptions, while the other one is concentrated on a specific \modifA{but
    unknown} support. Sample points from the latter are clustered based on a
  data-driven distance in a fully unsupervised approach. A theoretical grounding
  is provided through a link with the Christoffel function. Simulation results
  validate our approach and illustrate that it is an extension of a formerly
  proposed method.
\end{abstract}


\IEEEpeerreviewmaketitle

\section{Introduction}


In this paper, we consider multivariate data sets which come from probability
mixture models. Such models can describe data sets generated by systems
switching between different states, a situation which is likely to occur in many
circumstances (see e.g. \cite{rabiner89_tutor_hidden_markov_model_selec} for a
different context). On top of the switching nature of the system, we consider
that the recorded values stem from a linear transform of the signals of
interest. Contrary to many machine learning and neural networks based methods,
the context is here unsupervised or blind, which means that our method does not
require any training set.

On one side, mixture models have been extensively studied for a long time. A
classical point of view is to consider that the data generation is controlled by
a hidden variable taking values in a finite set. Many probability models have
been considered in this context, such as Markov models
\cite{rabiner89_tutor_hidden_markov_model_selec}, Markov fields or graphical
models \cite{koller09_probab_graph_model}. Recovering the hidden process is then
equivalent to a classification technique and this task is called unsupervised
whenever the model parameters are unknown.
Classification in presence of low probability events can also be seen as outlier
detection \cite{ducharlet22_lever_chris_darboux_kernel_onlin_outlier_detec}. Our
paper differs from the latter by making no difference between inliers and
outliers and by considering data with balanced proportions of the different
classes.

On the other side, observations resulting from a transform of the unknown data
is a commonly used model. Perturbation noise is generally added to the
degradation and the transform is often assumed linear
\cite{chouzenoux14_variab_metric_forwar_backw_algor,
  chaux07_variat_formul_frame_based_inver_probl}, although nonlinearity is a
more interesting but also much more challenging
situation~\cite{marmin21_spars_signal_recon_nonlin_model}. Contrary to many
recovery methods which require precise knowledge of the transform, we deal with
a blind context, assuming a linear but unknown transform, similarly to the
well-known Independent Component Analysis (ICA) or related source separation
techniques (see \cite{comon94:_indep_compon_analy, cardoso98:_blind,
  comon10_handbook_ica, hyvaerinen01_indep_compon_analy,
  simon01_separ_class_convol_mixtur} or
\cite{cruces23_minim_perim_criter_bound_compon_analy,
  weiss21_exact_algeb_blind_sourc_separ, saleh18_indep_compon_analy_based_non}
for recent works with a similar linear model).

This work combines both previously described contexts and is a significant
extension of \cite{castella-jasp13, rafi11_ica}. The novelties are:
\begin{enumerate}
\item we introduce a tool with strong theoretical \modifA{foundation} for
  unsupervised shape learning and detection of samples on a low dimensional
  nonlinear support \modifA{which is not known and not parameterized}.
\item we take advantage of an affine invariance property for combining this tool
  with a linear observation model.
\item compared with \cite{castella-jasp13, rafi11_ica}: (i) no model is required
  for the mixture components, (ii) no iterative procedure is required and (iii)
  multivariate data with dimension more than two can be dealt with.
\end{enumerate}

The problem and model are described in Section~\ref{sec:problem-statement}. An
intuitive presentation is given in Section~\ref{sec:nonl-dist-based} before
we provide a theoretical background in Section~\ref{sec:conn-with-christ}. The
overall procedure is given in Section~\ref{sec:appl-line-comp}. Simulations are
given in Section~\ref{sec:simulations} and Section~\ref{sec:conclusion}
concludes our work.

\vspace{-0.4em}
\section{Problem statement}
\label{sec:problem-statement}

\vspace{-0.2em}
\subsection{Linear mixture model}

Let us consider a data matrix $\bX=[\bx_1,\dots,\bx_T]\in\RR^{n\times T}$
corresponding to $T$ observed samples of an $n$-dimensional signal. The column
vectors $(\bx_t)_{t=1}^T$ are assumed to come from a linear mixture, that is,
there exists a fixed matrix $\bA\in\RR^{n\times n}$ which is assumed invertible
and another matrix $\bS=[\bs_1,\dots,\bs_T]\in\RR^{n\times T}$ such that
$\bX=\bA\bS$ or equivalently:
\begin{gather}
  \label{eq:linear_mixture}
   \bx_t=\bA\bs_t, \qquad \forall t\in\{1,\dots,T\}\,.
   \vspace{-0.4em}
 \end{gather}
 Both matrices $\bS$, the rows of which are called sources, and $\bA$, which is
 called mixing matrix, are unknown. The objective is to recover the sources in
 $\bS$ only from the recorded values in $\bX$. This task is equivalent to
 estimating an inverse $\widehat{\bB}$ of the mixing matrix. It is known that
 scaling and permutation ambiguities will necessarily remain in this blind
 context, which is similar to ICA \cite{comon94:_indep_compon_analy,
   cardoso98:_blind, comon10_handbook_ica, hyvaerinen01_indep_compon_analy}. In
 the latter method, a usual assumption is the non gaussianity and independence
 of the components in each random vector $(\bs_t)_{t=1}^T$. Here we consider on
 the contrary a model of dependent components.

\vspace{-0.4em}
\subsection{Probability mixture model}
\label{sec:prob-mixt-model}

We assume that the source samples $(\bs_t)_{t=1}^T$ are drawn according to a
probability mixture of two distributions $\PP_0$ and $\PP_1$.  Hence there
exists $\eta\in[0,1]$ such that the probability distribution of $\bs_{t}$ is
given by: 
\begin{gather}
  \label{eq:probmixture}
  \PP(\bs_t) = \eta\PP_0(\bs_t) + (1-\eta)\PP_1(\bs_t)\,, \quad \forall
  t\in\{1,\dots,T\}.
\end{gather}
An equivalent model consists in introducing a binary hidden (or latent) random
variable $r_{t}\in\{0,1\}$ such that $\PP(r_t=0) = \eta = 1- \PP(r_t=1)$.  The
vector $\bs_t$ can then be seen as the marginal of $(r_t,\bs_t)$, where
conditional distributions $\PP(\bs_t|r_t)$ are given by $\PP_0$ and $\PP_1$.
Due to the invertible linear relation (\ref{eq:linear_mixture}), the
distributions of $\bx_t$ and $\bs_t$ are deduced one from another, up to a
constant Jacobian term. Hence a strictly similar probability model holds for the
observed data $\bX$.  Finally, we also consider that, for different
$t\in\{1,\dots,T\}$, all variables are independent and identically distributed
(i.i.d.).  It follows that $\PP(\br,\bX) = \prod_{t=1}^T \PP(r_t)\PP(\bx_t|r_t)$
where the conditional distributions are respectively $\PP_0(\bx_t)$ and
$\PP_1(\bx_t)$. Although the same notation is used for both distributions of
$\bs_t$ and $\bx_t$, there should be no confusion: both models are similar and
only the latter will be involved in our method.

\vspace{-0.3em}
\subsection{Unsupervised classification problem}

Our methodology deals with situations where, due to the presence of $\PP_1$, the
classical assumption of ICA does not hold for $\PP$. However, the distribution
$\PP_0$, which is the same for all $t\in\{1,\dots,T\}$, is assumed to be such
that it satisfies the usual ICA requirements. In addition, $\PP_1$ is assumed to
be concentrated on a restricted nonlinear support, \modifA{which is
  unknown}. More precisely, the assumptions are:
\begin{hyp}
\item \label{hyp:2} $\PP_0(\bs_{t})$ is such that the components of $\bs_t$ are
  mutually independent and non Gaussian, except possibly one of them.
\item \label{hyp:1} The distribution $\PP_0$ is absolutely continuous with
  respect to the Lebesgue measure and the distribution $\PP_1$ is singular with
  support on an algebraic set of Lebesgue measure zero\footnote{This means in
    practice that the support of $\PP_1$ is the solution set of a finite number
    of polynomial equations.}.
\end{hyp}
To achieve reconstruction of $\bS$, a possible intermediate goal is to learn
\modifA{the unknown support of $\PP_1$} and classify the samples
$(\bx_t)_{t=1}^T$ according to whether $r_t=0$ or $1$.
Removing the points with $r_t=1$,
the remaining samples are drawn from $\PP_0$ and satisfy usual ICA
assumption. Then, it is possible to identify the inverse of $\bA$ by any
classical algorithm such as in \cite{cardoso93_blind_beamf_non_gauss_signal,
  hyvarinen99_fast_robus_fixed_point_algor, comon94:_indep_compon_analy}. This
basic idea has been introduced in \cite{castella-jasp13, rafi11_ica} and it has
been shown that unsupervised classification can be successful with a very
specific choice and knowledge of the corresponding model for $\PP_1$. Our method
here is more general.

\vspace{-0.3em}
\section{A nonlinear distance based classifier}
\label{sec:nonl-dist-based}

\subsection{Intuitive justification}

We will exploit the concentration of points of the distribution $\PP_1$ in a
specific region. Our method relies on quantifying how far a given sample deviate
from it. For any vector $\bx=(x_1,\dots,x_n)\tr\in\RR^n$, a classical squared
distance to the point cloud given by the samples in $\bX$ is given by the
quantity $(\bx-\bmu)\tr\bSigma^{-1}(\bx-\bmu)$, where, writing $\bone$ an
all-one column vector of size $T$, $\bmu = \frac{1}{T}\bX\bone$ and
$\bSigma=\frac{1}{T}\bX\bX\tr-\bmu\bmu\tr$ are the empirical mean and covariance
matrices.  Points equally far from the mean lie on an ellipsoid defined by the
covariance matrix.  Alternatively, one can include the constant $1$ in the data
feature space and introduce the extended covariance matrix
\begin{gather}
  \label{eq:cov_extended}
  \widetilde{\bSigma}
  = \frac{1}{T}\sum_{t=1}^T
  \begin{bmatrix}
    1 \\ \bx_t
  \end{bmatrix}
  \begin{bmatrix}
    1 & \bx_t\tr
  \end{bmatrix} \,.
  \vspace{-0.3em}
\end{gather}
Using Schur complement as in \cite{lasserre16_sortin_out_typic_with_inver}, we
obtain the same distance criterion up to a constant:
\begin{gather}
  \label{eq:leverage2}
  \begin{bmatrix}
    1 & \bx\tr
  \end{bmatrix}
  \widetilde{\bSigma}^{-1}
  \begin{bmatrix}
    1 \\ \bx
  \end{bmatrix}
  =
  (\bx-\bmu)\tr\bSigma^{-1}(\bx-\bmu) + 1 \,.
\end{gather}
The above notion, which is linked to an implicit Gaussian assumption, appears
under the name of leverage-score or Mahalanobis distance
\cite{bishop06_patter_recog_machin_learn}.  In our context, we assume that the
data drawn according to $\PP_1$ is concentrated in the neighborhood of a
lower dimensional subspace defined by nonlinear equations. A natural idea in a
nonlinear context consists in further extending the data feature space by
including additional monomials in a spirit similar to Taylor expansions or
Volterra filters.

\vspace{-0.3em}
\subsection{Proposed method}
\label{sec:proposed-method}

For any order $d\in\NN$, denote by $[\bx]_d$ a column vector containing a basis
of all polynomials in $\bx$ with maximal degree $d$. In practice, we included in
$[\bx]_d$ all monomials of degree less than or equal to $d$. To determine the
points corresponding to $r_t=1$, we propose to use this extended vector and
compute for all $t\in\{1,\dots,T\}$ a score $\theta_t$ based on Equations
(\ref{eq:cov_extended}) and (\ref{eq:leverage2}). This score is then compared to
a threshold value $\overline{\theta}$, the choice of which will be discussed
later. The procedure for finding an estimate $\hat{r}_t$ of $r_t$ hence consists
of the steps given in Alg. \ref{alg:classif}. 
\vspace{-0.3em}
\begin{algorithm}
  \caption{Classification method}
  \label{alg:classif}
  \textbf{Input:} Data matrix $\bX=(\bx_t)_{t=1}^T$, threshold value $\overline{\theta}$.
  \begin{enumerate}
  \item Compute the extended empirical covariance matrix:
    \begin{gather}
      \label{eq:empir_mom_matrix}
      \widehat{\bM}_d = \frac{1}{T} \sum_{t=1}^T [\bx_t]_d[\bx_t]_d\tr \,.
      \vspace{-0.3em}
    \end{gather}
  \item For $t=1,\dots,T$, compute
    \begin{gather}
      \label{eq:thetat}
      \theta_{t} = [\bx_t]^T\big(\widehat{\bM}_d\big)^{-1}[\bx_t] \,.
      \vspace{-0.3em}
    \end{gather}
  \item Set $\hat{r}_t= \begin{cases}
      0 & \text{if } \theta_t>\overline{\theta} \,,\\
      1 & \text{if } \theta_t<\overline{\theta} \,.
    \end{cases}$
  \end{enumerate}
  \textbf{Output:} Estimated classification $\hat{\br}=(\hat{r}_t)_{t=1}^T$.
\end{algorithm}
\vspace{-0.3em}

\section{Connection with Christoffel-Darboux kernel}
\label{sec:conn-with-christ}

A theoretical justification of our method is provided by a link with the
Christoffel function and the Christoffel-Darboux kernel. They have been
known for long and are classical tools in interpolation and approximation with a
close link to orthogonal polynomials. Their usefulness and relevance for data
analysis tasks have been recently recognized
\cite{lasserre16_sortin_out_typic_with_inver,
  lasserre19_empir_chris_funct_with_applic_data_analy,
  lasserre22_chris_darboux_kernel_data_analy}. A major asset of the Christoffel
function is its ability to encode information about the shape of a
distribution and, more importantly for us, it can detect the presence of a
singular continuous
component~\cite{pauwels21_data_analy_from_empir_momen_chris_funct,
  korda20_data_driven_spect_analy_koopm_operat}.

\subsection{Definitions and properties}
\label{sec:defin-prop}

Consider the probability distribution $\PP(\bx)$ on the observed variables and
assume that it is supported on a compact set 
$\bK\subset \RR^n$. The associated moment matrix is by definition:
\begin{gather}
  \label{eq:mom_matrix}
  \bM_d^{\PP} = \int_{\bK} [\bx]_d[\bx]_{d}\tr \dint\PP(\bx) \,,
\end{gather}
where the integral is taken component-wise.
Since any polynomial with degree less than $d$ can be written
$p(\bx)=\bp\tr[\bx]_d$ with $\bp$ the corresponding coefficients vector in the
basis $[\bx]_d$, one can see that
\begin{gather*}
  \bp\tr\bM_d^{\PP}\bp = \int_{\bK}p(\bx)^2\dint\PP(\bx) \,.
\end{gather*}
Therefore, $\bM_d^{\PP}$ is symmetric positive semi-definite. It is also
positive definite under a non degeneracy condition which is satisfied for
absolutely continuous measures \cite{lasserre16_sortin_out_typic_with_inver,
  lasserre22_chris_darboux_kernel_data_analy}. In our context, the presence of
$\PP_0$ with assumption \ref{hyp:1} ensures that $\bM_d^{\PP}$ is non
singular. Writing $\by=(y_1,\dots,y_n)\tr$, the Christoffel-Darboux kernel
associated to $\PP$ can then be defined by
\begin{gather*}
  \kappa_d^{\PP}(\bx,\by) = [\bx]_d\tr\left(\bM_d^{\PP}\right)^{-1}[\by]_d \,.
\end{gather*}
For any $\bz=(z_1,\dots,z_n)\tr$, another quantity of interest, referred to as
the Christoffel function, is given by
$C_d^{\PP}(\bz)=\frac{1}{\kappa_d^{\PP}(\bz,\bz)}$. It can be equivalently
defined based on the following variational formula, where the minimization is
with respect to polynomials with degree less than $d$ and taking value $1$ at
$\bz$ (see \cite{lasserre16_sortin_out_typic_with_inver,
  lasserre22_chris_darboux_kernel_data_analy} for details):
\begin{gather}
  \label{eq:CF_def_variational}
  C_d^{\PP}(\bz) = \min_{p\in\RR[\bx]_d, \,p(\bz)=1} \int p(\bx)^2 \dint\PP(\bx) \,.
\end{gather}
From the above formula, one can understand that the shape of regions with high
probability mass can be captured.

\subsection{Empirical Christoffel function}

In our practical setting, our method relies on the matrix $\widehat{\bM}_d$ from
Equation (\ref{eq:empir_mom_matrix}). Since the sample values $(\bx_t)_{t=1}^T$
stored in $\bX$ are i.i.d. and follow the distribution $\PP$, $\widehat{\bM}_d$ in
(\ref{eq:empir_mom_matrix}) is the empirical estimate of the matrix
$\bM_d^{\PP}$ from Equation (\ref{eq:mom_matrix}). Importantly, as noted in
\cite{lasserre19_empir_chris_funct_with_applic_data_analy}, because
$\bM_d^{\PP}$ is non singular, it holds for $T$ large enough that $\widehat{\bM}_d$
is almost surely invertible.  As a consequence, the score $\theta_t$ computed in
our method is the empirical estimate at $\bx_t$ of
$\bx\mapsto\kappa_d^{\PP}(\bx,\bx)$, which is the inverse of the Christoffel
function. Note that it has been shown in
\cite{lasserre19_empir_chris_funct_with_applic_data_analy} that the empirical
Christoffel function converges almost surely and uniformly in $\bx$ to
$C^{\PP}_d(\bx)$ for large~$T$.

\subsection{Support and shape detection}
\label{sec:supp-shape-detect}

\subsubsection{Case of a singular support}

Corresponding to the fact that $\PP_1$ is concentrated on a specific set, we
made Assumption \ref{hyp:1} concerning the decomposition of $\PP$ in
(\ref{eq:probmixture}) as a probability mixture.
No additional assumption is made and in particular, nothing more is known about
the support of $\PP_1$. Contrary to \cite{castella-jasp13, rafi11_ica}, no model
is introduced for $\PP_1$. The task of learning $\br$ is exclusively based on
unsupervised identification of the support of $\PP_1$.

For any measure such as $\PP_1$ with singular support, the definition of the
Christoffel-Darboux kernel requires attention because of the singularity of
$\bM_d^{\PP_1}$.  Fortunately the variational definition in
(\ref{eq:CF_def_variational}) remains valid.  From the latter, because any
polynomial with $p(\bz)=1$ is positive in a small neighborhood of $\bz$, we have
$C_d^{\PP_1}(\bz)>0$ for any $\bz$ in the support of $\PP_1$.
More importantly for us, it has been proven that outside the support of $\PP_1$,
the inverse Christoffel function $\bx\mapsto\kappa_d^{\PP_1}(\bx,\bx)$ on which
our method is based grows (in $d$) at an exponentially fast rate and hence the
Christoffel function goes to zero
\cite{lasserre19_empir_chris_funct_with_applic_data_analy}. Knowing the
Christoffel function associated to a measure therefore helps identifying its
support.

\subsubsection{Threshold value} The previous elements justify to consider points
with large values of the inverse Christoffel function as an estimation for
points \modifA{outside} the singular component $\PP_1$ of $\PP$: this is precisely what is
done in our method by computing $\theta_t$ for each $t\in\{1,\dots,T\}$ and
comparing it to the threshold $\overline{\theta}$. More precisely, based on
\cite{lasserre19_empir_chris_funct_with_applic_data_analy}, $\overline{\theta}$
should be proportional to the binomial coefficient
$\binom{n+d}{n}=\frac{(n+d)!}{n!d!}$. This is confirmed in
\cite{ducharlet22_lever_chris_darboux_kernel_onlin_outlier_detec} and we
precisely choose $\overline{\theta} = \eta \binom{n+d}{n}$ in our experiments.

\section{Application to linear component analysis}
\label{sec:appl-line-comp}

We come back to our initial objective of obtaining a linear decomposition
similar to the ICA model in (\ref{eq:linear_mixture}). Therefore, it is natural
to ask how the tools introduced previously behave under linear transformation of
the data.

\subsection{Affine invariance}
Let us consider an invertible matrix $\bB\in\RR^{n\times n}$. Given the
probability distribution $\PP$ on $\bx$, this matrix induces a probability
distribution denoted $\PP_{\bB}$ for the corresponding variable $\bz=\bB\bx$.
It has been proven (see e.g. \cite{lasserre16_sortin_out_typic_with_inver,
  lasserre19_empir_chris_funct_with_applic_data_analy}) that the
Christoffel-Darboux function satisfies an invariance property by any invertible
affine transform and in particular:
\begin{gather*}
  \kappa_d^{\PP}(\bx,\bx) = \kappa_d^{\PP_{\bB}}(\bB\bx,\bB\bx) 
\end{gather*}
As a consequence, we have the following proposition:
\begin{prop}
  For a given realization of $(\bs_{t})_{t=1}^T$ and for any $(\bx_{t})_{t=1}^T$
  such that (\ref{eq:linear_mixture}) holds with invertible $\bA$, the scores
  $(\theta_t)_{t=1}^T$ defined in (\ref{eq:empir_mom_matrix})-(\ref{eq:thetat})
  take values independent of $\bA$.
\end{prop}
This property is of high importance in our method: contrary to
\cite{castella-jasp13, rafi11_ica}, we will not use any iterative procedure, but
the values of $(\theta_t)_{t=1}^T$ will be computed only once based on
$(\bx_t)_{t=1}^T$. These values are identical to the values that would have been
computed based on $(\bs_t)_{t=1}^T$.

\subsection{Method for linear decomposition}

In our context of linear mixture, thanks to the above affine invariance, the
Christoffel-Darboux function is a particularly well suited tool for classifying
points from components $\PP_0$ or $\PP_1$. Relying on an existing ICA algorithm
denoted $\mathtt{ICA}(.)$ that returns the inverse of the mixing matrix such as
CoM2 \cite{comon94:_indep_compon_analy}, FastICA
\cite{hyvarinen99_fast_robus_fixed_point_algor}, JADE
\cite{cardoso93_blind_beamf_non_gauss_signal,
  cardoso99_high_order_contr_indep_compon_analy}, the proposed global procedure
also performs a linear decomposition as in (\ref{eq:linear_mixture}). We sum it
up in Alg.~\ref{alg:lin_decomp}.
\begin{algorithm}
  \caption{Linear decomposition}
  \label{alg:lin_decomp}
  \textbf{Input:} Data matrix $\bX=(\bx_t)_{t=1}^{T}$, algorithm
  $\mathtt{ICA}(.)$
  \begin{itemize}
  \item Perform steps 1) to 3) as in Section \ref{sec:proposed-method},
    Alg.~\ref{alg:classif}.
  \item Define $\widehat{\bX}_0$ the submatrix of $\bX$ with columns indexed by
    $\{t=1,\dots,T \,|\, \hat{r}_t=0\}$.
  \item Perform $\widehat{\bB} =
    \mathtt{ICA}(\widehat{\bX}_0)$. 
  \end{itemize}
  \textbf{Output:} Estimated linear components $\widehat{\bS}=\widehat{\bB}\bX$.
\end{algorithm}

\section{Simulations}
\label{sec:simulations}

\vspace{-0.2em}
\subsection{Experimental setup}
\label{sec:experimental-setup}

We have tested our method on synthetic data where the sources in $\bS$ were
randomly drawn according to (\ref{eq:probmixture}). The distribution $\PP_0$
satisfied (\ref{hyp:2}-\ref{hyp:1}) with components uniformly distributed,
centered and unit variance. The different choices for $\PP_1$ are detailed in
the next section.
The matrix $\bA$ has been systematically randomly drawn with i.i.d. Gaussian
entries and our ICA algorithm was CoM2 \cite{comon94:_indep_compon_analy}. To
quantify the success of our method, we eliminated the inherent ambiguities of
ICA and considered the average mean square error (MSE) on the components of the
recovered $(\bs_t)_{t=1}^T$.
In addition the probability of correctly estimating $(r_t)_{t=1}^T$, has been
computed, given by $\Upsilon = \frac{\#\{t\,|\,\hat{r}_t=r_t\}}{T}$ where $\#$
is the \modifB{cardinality} of the set.  For comparison, \modifA{we
  considered
  the result of ICA} applied directly on the data $\bX$, hence ignoring the
probability model assumed in Section~\ref{sec:prob-mixt-model}.  We also
considered the ideal supervised case with known true values of $(r_t)_{t=1}^T$,
keeping the samples with $r_t=0$ as an input for the ICA algorithm. All
presented results are
mean values over 1000 Monte-Carlo realizations, after discarding the bottom/top
1\% values.

\vspace{-0.2em}
\subsection{Simulation results}
\subsubsection{Comparison with \cite{castella-jasp13}}
\label{sec:comp-with-citec}
The dependent sources given by Example~1 in \cite{castella-jasp13} satisfy
conditions (\ref{hyp:2}-\ref{hyp:1}) \modifA{and our method is indeed successful
  in separating them with a computational complexity that is reduced compared to
  the iterative procedure in \cite{castella-jasp13} (see
  Table~\ref{table:ex1}).}  As shown next, our method goes beyond this very
specific case where $n=2$.
\begin{table}[htbp]
  \begin{center}
     \setlength\tabcolsep{0pt}
    \begin{tabular}{p{0.16\linewidth}|@{\hspace{0.02\linewidth}}p{0.38\linewidth}p{0.15\linewidth}p{0.15\linewidth}p{0.15\linewidth}}
      \toprule
      \modifA{Runtime} & Method & $\eta=0.2$ & $\eta=0.4$ & $\eta=0.6$ \\
      \midrule
      \modifA{1.8 ms} & Ignore $\PP_1$
      & 0.5849 & 0.5838 & 0.2493 \\
      \modifA{33.6 ms} & \modifA{Method in \cite{castella-jasp13}} & \modifA{0.5848} & \modifA{0.1953} & \modifA{0.0072} \\
      \modifA{5.1 ms} & Proposed  (order $d=6$)
      & \textbf{0.1232} & \textbf{0.0763} & \textbf{0.0389} \\
      \modifA{1.2 ms} & \modifB{Known} $\br$ 
      & 0.0011 & 0.0006 & 0.0003 \\
      \bottomrule
    \end{tabular}
    \end{center}
    \caption{MSE results on sources from \cite[Example~1]{castella-jasp13}
      (T=2000 samples) \modifA{and comparison of runtime with $\eta=0.6$}.}
    \label{table:ex1}
\end{table}

\vspace{-0.5em}
\subsubsection{Case of 3 sources}
\label{sec:case-3-sources}
\begin{figure}[htbp]
  \includegraphics[width=0.99\linewidth]{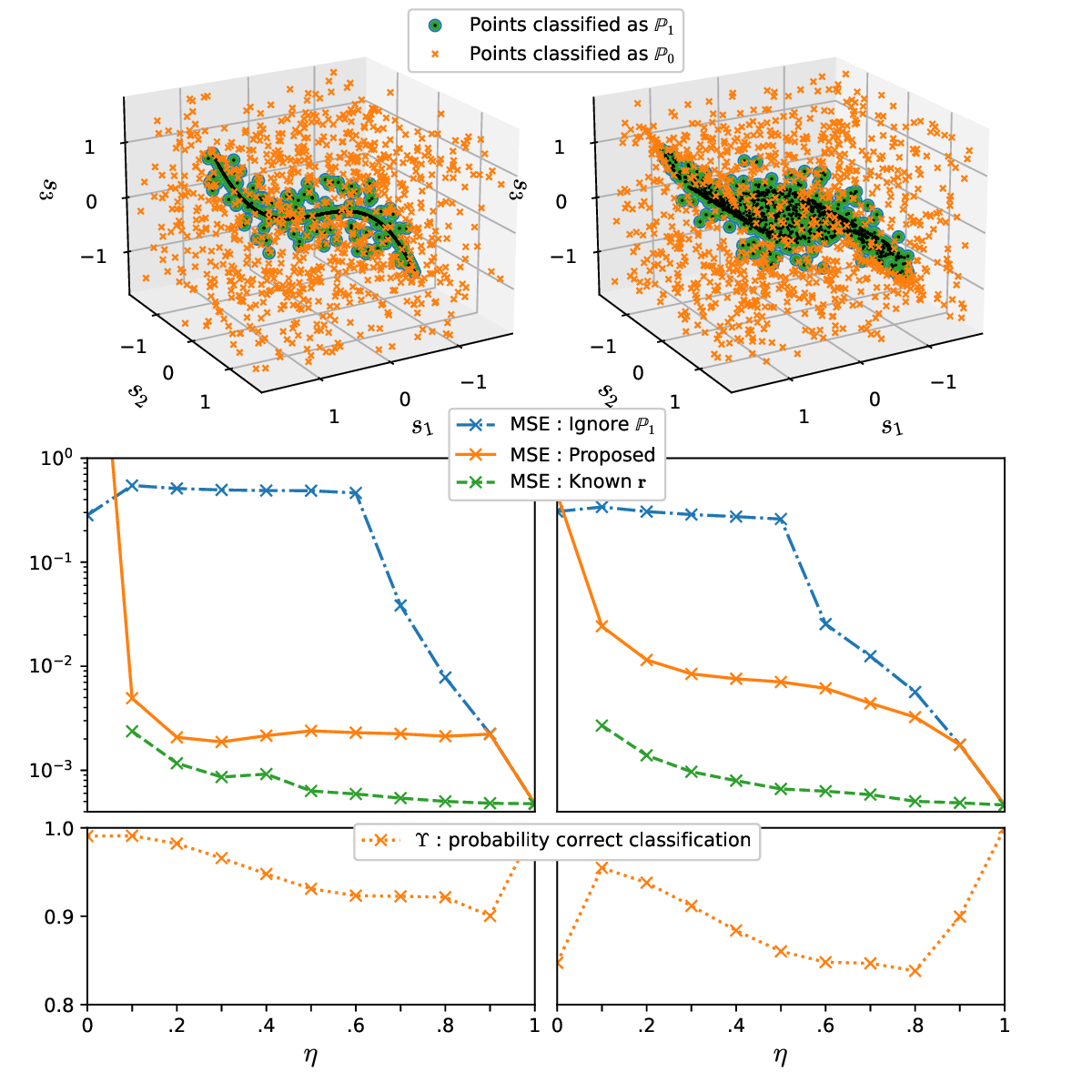}
  \vspace{-0.5em}
  \caption[]{Sources from Sec.~\ref{sec:case-3-sources}:
    typical classification and corresponding linear component analysis results
    ($T=2000$ samples, order $d=6$).
  }
  \label{fig:ex3c}
  \vspace*{-1em}
\end{figure}

As an extension, we considered $n=3$ sources where $\PP_1$ is given as follows:
$s_1$ uniform on $[-\beta,\beta]$, $s_2$ uniform on $[-\gamma,\gamma]$ and
$s_3=\frac{1}{\beta^2}s_1^3$.  We took $\beta=3/2$ and a typical realization of
such sources is given on the top row of Figure~\ref{fig:ex3c}, for $\gamma=0$
(left) and $\gamma=1$ (right): to show the success of our unsupervised
classification, the points classified with $\hat{r}_t=1$ are plotted in green
and the shape of the hypersurface appears clearly. Correspondingly, the values
of the MSE (middle) and of $\Upsilon$ are plotted depending on $\eta$.  Our
method has good performance \modifA{and shows better results with data
  concentrated on a low dimensional subset.}

\subsubsection{Simultaneously vanishing sources}
\label{sec:simult-vanish-sourc}
For a more concrete example, we considered a scenario where two sources are
simultaneously switched off. This might occur in different applications.
We took $n=5$ random independent uniformly distributed sources
when $r_t=0$, whereas we set components 4 and 5 of $\bs_t$ to zero when $r_t=1$.
Table~\ref{table:ex5} shows the success of our method.
\modifB{The best results were obtained experimentally with the order $d=6$}, which seems a
compromise between good modeling capabilities and numerical stability.
\begin{table}[htbp]
  \begin{center}
     \setlength\tabcolsep{0pt}
    \begin{tabular}{p{0.40\linewidth}p{0.15\linewidth}p{0.15\linewidth}p{0.15\linewidth}p{0.15\linewidth}}
      \toprule
      Method
      & $\eta=0.2$ & $\eta=0.4$ & $\eta=0.6$ & $\eta=0.8$ \\
      \midrule
      Ignore $\PP_1$
      & 0.2239 & 0.1994 & 0.1988 & 0.0056 \\
      Proposed  (order $d=2$)
      & 0.0647 & 0.0140 & 0.0089 & 0.0056 \\
      Proposed  (order $d=4$)
      & 0.0103  & 0.0067 & 0.0049 & 0.0038 \\
      Proposed  (order $d=6$)
      & \textbf{0.0058}  & \textbf{0.0044} & \textbf{0.0038} & \textbf{0.0028} \\
      Proposed  (order $d=8$)
      & 0.0042 & 0.0042 & 0.0048 & 0.0035 \\
      \modifB{Known} $\br$
      & 0.0034 & 0.0018 & 0.0013 & 0.0011 \\
      \bottomrule
    \end{tabular}
  \end{center}
  \caption{MSE results on sources from Sec.\ref{sec:simult-vanish-sourc} (T=2000 samples).}
  \label{table:ex5}
  \vspace{-0.5em}
\end{table}

\vspace{-2em}
\section{Conclusion}
\label{sec:conclusion}

We have considered an extension of ICA for data switching between two
probability distributions, only one of which satisfies the ICA assumptions. For
such a case, we proposed an intuitively simple and theoretically grounded method
for performing a linear decomposition in a blind context. An unsupervised
selection of data samples in accordance with the ICA assumptions is performed by
identifying points clustered in a restricted region through the use of the
Christoffel function. Due to its affine invariance, this tool is particularly
well suited for this linear superposition context. Simulations show the interest
and good performance of the approach for different examples and models,
including cases where a previously proposed approach is not applicable.

\bibliographystyle{\pathStyleFiles/IEEEtran}
\bibliography{\pathBibFilesA/BiblioLoc}

\ifextended
\clearpage
\appendix
\modifA{We gather in this Appendix some supplemental material and results that
  may be \modifB{useful}.}

\section*{Notation clarification (example))}

\modifB{In this paper, the notation $[\bx]_d$ denotes a vector containing a
  basis of polynomials in $\bx$ with maximal degree $d$. For example, if the
  monomial basis is considered, this yields with $n=2$ and $d=2$, $d=3$
  respectively:
  \begin{gather*}
    [\bx]_2 =
    \begin{bmatrix}
      1 \\ x_1 \\ x_2 \\ x_1^2 \\ x_1x_2 \\ x_2^2
    \end{bmatrix}
    \quad
    [\bx]_3 =
    \begin{bmatrix}
      1 \\ x_1 \\ x_2 \\ x_1^2 \\ x_1x_2 \\ x_2^2 \\ x_1^3 \\ x_1^2 x_2 \\ x_1x_2^2 \\ x_2^3      
    \end{bmatrix}
  \end{gather*}
  It is known that the vector space of polynomials of degree at most $d$ in
  $\bx=(x_1,\dots,x_n)$ has dimension $\binom{n+d}{n}=\frac{(n+d)!}{n!d!}$,
  which is therefore the size of $[\bx]_d$.}

\section*{Additional details on Section \ref{sec:comp-with-citec}}
\modifA{The plot of the complete simulation results corresponding to
  Table~\ref{table:ex1} is given below. For small and big values of $\eta$ (that
  is $\eta<0.5$ or $\eta\geq 0.8$), one can see that the proposed method
  performs better than the method in \cite{castella-jasp13}, although the
  generated data is more favorable to the method in \cite{castella-jasp13}. Note
  also that for values of $\eta\geq0.8$, the perturbation introduced by $\PP_1$
  is small enough and better or equivalent results are obtained by just applying
  ICA on the perturbed data.}

\includegraphics[width=7cm]{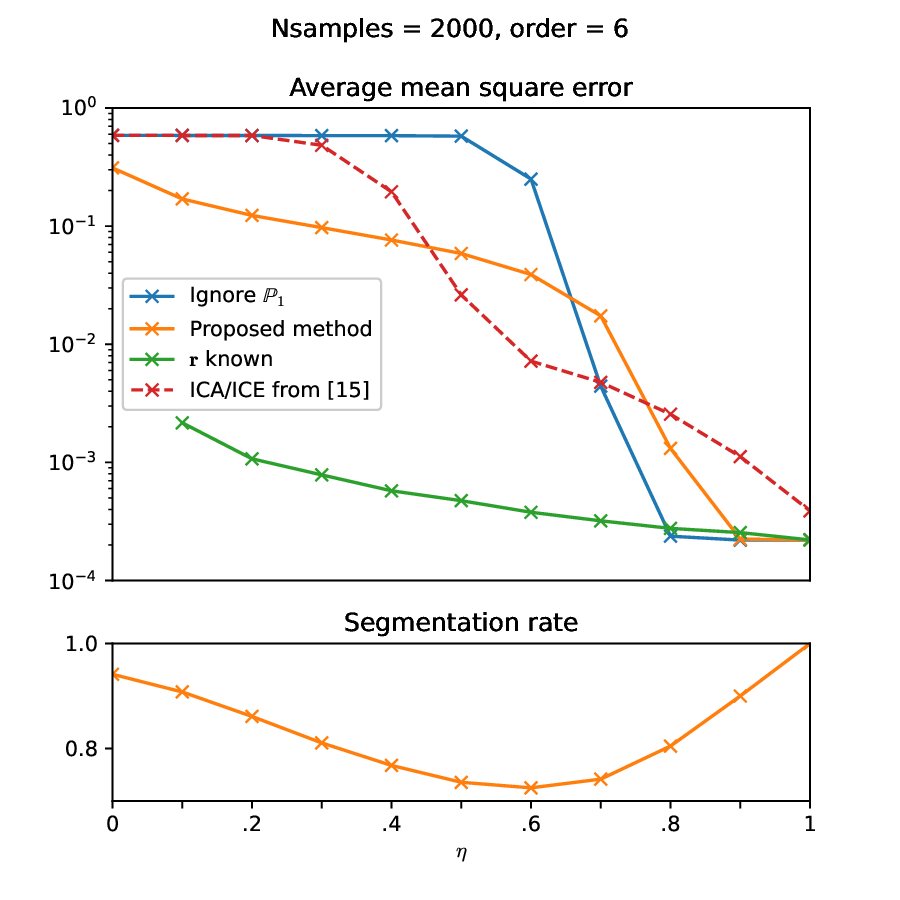}

\section*{Additional details on Section \ref{sec:case-3-sources}}

\modifA{Below are results showing more precisely how our method behaves for $\eta$
between $0.8$ and $1$ for the corresponding sources of
Figure~\ref{fig:ex3c}. One can see that above a threshold value of $\eta$
(approximately $0.85$), our method yields results corresponding to the
classification $\hat{r}_t=0$ for all $t$.}

\includegraphics[width=\linewidth]{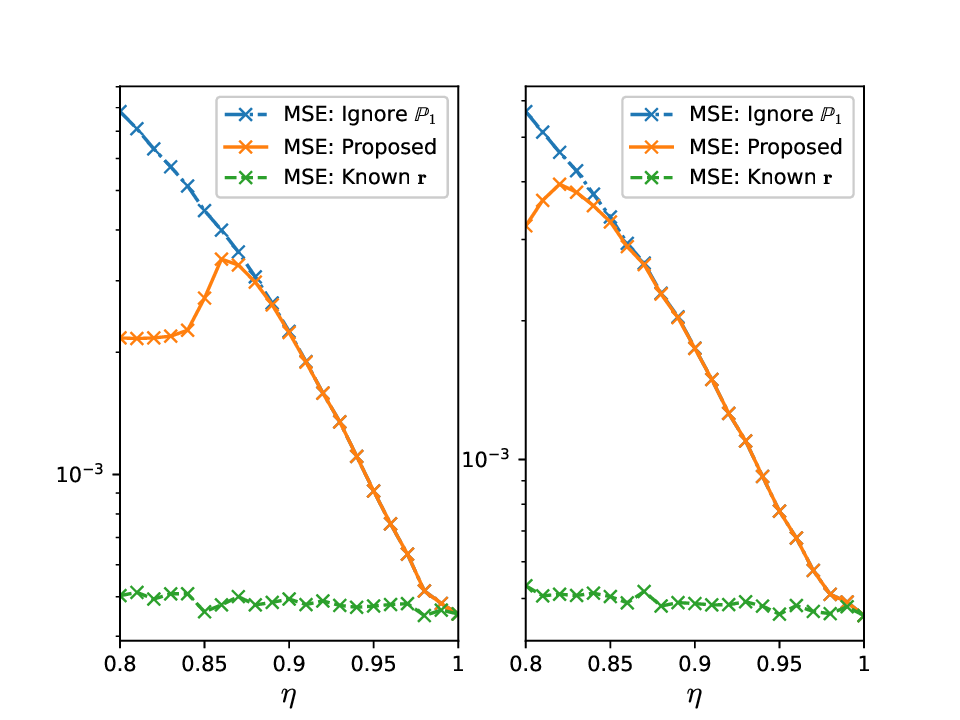}

\section*{Comment on the choice of $d$}

\modifB{As mentioned above, the size of $[\bx]_d$ is $\binom{n+d}{d}$. It is
  also the size of the square matrices $\bM_d$ and $\widehat{\bM}_d$. The
  corresponding values are given below depending on $n$ and $d$:\hfill\mbox{}
  \begin{center}
    \begin{tabular}{c|ccccc}
      & $d=1$ & $d=2$ & $d=4$ & $d=6$ & $d=8$ \\
      \midrule
      $n=2$ & 3 & 6  & 15  & 28   & 45 \\
      $n=3$ & 4 & 10 & 35  & 84   & 165 \\
      $n=5$ & 6 & 21 & 126 & 462  & 1287 \\
      $n=8$ & 9 & 45 & 495 & 3003 & 12870 \\
      \bottomrule
    \end{tabular}
  \end{center}
  The complexity of our method given in Alg.~\ref{alg:classif} is linked to the
  inversion of $\widehat{\bM}_d$ and hence to its size $\binom{n+d}{d}$ as given
  above. Note that when the inversion of $\widehat{\bM}_d$ is too costly, it is
  possible to solve a quadratic optimization problem for each evaluation of
  $\theta_t$ in Equation (\ref{eq:thetat}) (see
  \cite{lasserre22_chris_darboux_kernel_data_analy}).}

\modifB{Finally, let us mention that choosing the hyperparameter $d$ requires a
  compromise: greater values of $d$ indeed provide better modeling
  capabilities. However, we observed badly conditioned moment matrices for
  values above $d\geq 8$. This phenomenon seems more limiting than computational
  load difficulties.}
\else\fi

\end{document}

%% file: defSPL.tex
\usepackage[utf8]{inputenc}
\usepackage{amsmath}
\usepackage{dsfont}
\usepackage{amssymb} 
\usepackage[mathscr]{eucal}
\usepackage{xcolor}
\usepackage{multirow}           %
\usepackage{booktabs}           
\ifCLASSOPTIONcompsoc
\usepackage[caption=false,font=normalsize,labelfont=sf,textfont=sf]{subfig}
\else
\usepackage[caption=false,font=footnotesize]{subfig}
\fi
\usepackage{algorithm} \usepackage{algorithmic}
\usepackage{enumitem}           

\ifCLASSINFOpdf
   \usepackage[pdftex]{graphicx}
   \graphicspath{{./fig_eps/}{./src_img/}}
  \DeclareGraphicsExtensions{.pdf,.jpeg,.png}
\else
  \usepackage[dvips]{graphicx}
  \graphicspath{{./fig_eps/}{./src_img/}}
  \DeclareGraphicsExtensions{.eps}
\fi

%% file: defCdes.tex
\newcounter{hypie} \newcounter{hyp}
\newenvironment{hyp}
     {\begin{list}{\textsf{A\arabic{hyp}.}}
                  { \usecounter{hyp}   \setcounter{hyp}{\value{hypie}}
                   \setlength{\listparindent}{0pt}
                   \setlength{\labelwidth}{0pt}
                   \setlength{\leftmargin}{10pt}
                   } }
     {\end{list} \setcounter{hypie}{\value{hyp}}}

\newcounter{rqueie} \newcounter{rque}

\newtheorem{prop}{Proposition}

\def\chooseColor#1{%
  \ifcase#1\or red\or green\or blue\or cyan\or magenta\or gray\fi}

\newcommand{\temp}[1]{{\color{red} *** \textsc{ #1} ***}}
\newcommand{\modifA}[1]{\textcolor{red}{#1}}
\newcommand{\modifB}[1]{\textcolor{magenta}{#1}}

\newcommand{\NN}{\mathbb{N}}
\newcommand{\PP}{\mathbb{P}}

\newcommand{\RR}{\mathbb{R}}

\newcommand{\dint}{\mathop{}\!\mathrm{d}}

\newcommand{\tr}{{^\top}} 

    \DeclareFontFamily{U}{wncy}{}
    \DeclareFontShape{U}{wncy}{m}{n}{<->wncyr10}{}
    \DeclareSymbolFont{mcy}{U}{wncy}{m}{n}
    \DeclareMathSymbol{\Sh}{\mathord}{mcy}{"58}

\def\qed{\ifmmode\hbox{\hfill\sqb}\else{\ifhmode\unskip\fi%
\nobreak\hfil
\penalty50\hskip1em\null\nobreak\hfil$\blacksquare$
\parfillskip=0pt\finalhyphendemerits=0\endgraf}\fi}


%% file: defGras.tex
\newcommand{\bone}{\mathbf{1}}

\newcommand{\bA}{\mathbf{A}}
\newcommand{\bB}{\mathbf{B}}

\newcommand{\bK}{\mathbf{K}}

\newcommand{\bM}{\mathbf{M}}

\newcommand{\bS}{\mathbf{S}}

\newcommand{\bX}{\mathbf{X}}

\newcommand{\bp}{\mathbf{p}}

\newcommand{\br}{\mathbf{r}}
\newcommand{\bs}{\mathbf{s}}

\newcommand{\bx}{\mathbf{x}}
\newcommand{\by}{\mathbf{y}}
\newcommand{\bz}{\mathbf{z}}

\newcommand{\bmu}{\boldsymbol{\mu}}

\newcommand{\bSigma}{\mathbf{\Sigma}}